\documentclass[a4paper,12pt]{article}
\usepackage[pdftex]{graphicx}
\usepackage[T1,T2A]{fontenc}
\usepackage[cp1251]{inputenc}
\usepackage[russian]{babel}
\usepackage{amssymb}
\usepackage{amsmath}
\usepackage{setspace}

\addto\captionsrussian{}
\usepackage[T2A]{fontenc}

\begin{document}

\def\figurename{Fig.}
\def\abstractname{Abstract}
\def\bibname{References}

\setcounter{page}{1}

\begin{center}
{\large
{\bf 
Magnetocaloric Effect in Two-Dimensional Diluted Ising Model: Appearance of Frustrations in the Ground State
} 
}
\vskip0.5\baselineskip{
\bf 
A.V. Shadrin$^{a,*}$,
V.A. Ulitko$^a$,
Y.D. Panov$^a$
}
\vskip0.5\baselineskip{
$^a$Ural Federal University Named after the First President of Russia B.N. Yeltsin, Yekaterinburg, Russia

$^*$ e-mail: shadrin.anton@urfu.ru
}

\end{center}

\begin{abstract}

The magnetocaloric effect in a two-dimensional Ising model is considered for different ratios
between parameters of inter-site repulsion of nonmagnetic impurities and exchange coupling. Classical
Monte Carlo simulations on a square lattice show that in case of weak coupling and at sufficiently high concentrations of nonmagnetic impurities the long-range ferromagnetic ordering breaks down to give isolated
spin clusters in the ground state of the system. This leads to appearance of a paramagnetic response in the
system at the zero temperature and nonzero entropy of the ground state. The feasibility to detect frustration
of ground state using data on the magnetic entropy variation is discussed.

\end{abstract}

\section{Introduction}

The magnetocaloric effect (MCE) consists in
release or absorption of heat as a result of variation of
an external magnetic field applied to a material. Initially, the effect was used to reach temperatures below
1 K, but, since materials exhibiting MCE at near room
temperature have been discovered, magnetic cooling
has become an area of very active research. The MCE
in frustrated and low-dimensional systems attracts a
special interest \cite{Zhitomirsky2003,Zhitomirsky2004,Honecker2006,Schmidt2007}. In Ising-type two-dimensional
systems, the dependence of MCE on the parameter of
magnetoelastic interaction for a square lattice was
investigated in \cite{Amaral2014}, while the effects of shape and size
of geometrically frustrated Ising spin clusters in a triangular lattice have on the MCE were addressed in \cite{Zukovic2015}.
In this work, we consider a two-dimensional Ising
system with a fixed concentration of mobile nonmagnetic charged impurities. The dilute Ising model is one
key model \cite{Katsura1965,Blume1971} in the theory of magnetic systems
with quenched or annealed disorder as well as in the
thermodynamic theory of binary alloys and mixtures
of classical and quantum liquids.
To describe our system, we use the pseudo-spin
formalism ($S = 1$) in which for a given lattice site the
states with pseudo-spin projections $S_z = \pm 1$ correspond to two magnetic states with spin projections $s_z = \pm 1/2$, while the state with $S_z = 0$ corresponds to a
charged nonmagnetic impurity.
Write the Hamiltonian as follows:

\begin{equation}\label{MainHamiltonian}
 H = -\tilde{J} \sum\limits_{\langle ij \rangle} S_{zi} S_{zj} + V \sum\limits_{\langle ij \rangle} P_{0i} P_{0j} - h \sum\limits_{i} S_{zi},
\end{equation}
where $S_{zi}$ is the $z$-projection of pseudo-spin operator
on a site, $P_{0i} = 1 - S_{zi}^2$ is the projection operator on
state $S_{zi} = 0$, $\tilde{J} = J s^2$, $J$ is an exchange integral,  $s=1/2$, $V$ is the inter-site interaction between impurities, $h$ is
an external magnetic field, $\langle ij \rangle$ -- are the nearest neighbors, and the sum is taken over all sites of the twodimensional square lattice. The concentration n of
charged nonmagnetic impurities is fixed so that $nN = \sum\limits_i P_{0i} = const$ , where $N$ is the number of lattice sites.

\begin{figure}[h]
   \centering
   \includegraphics[width=0.95\linewidth]{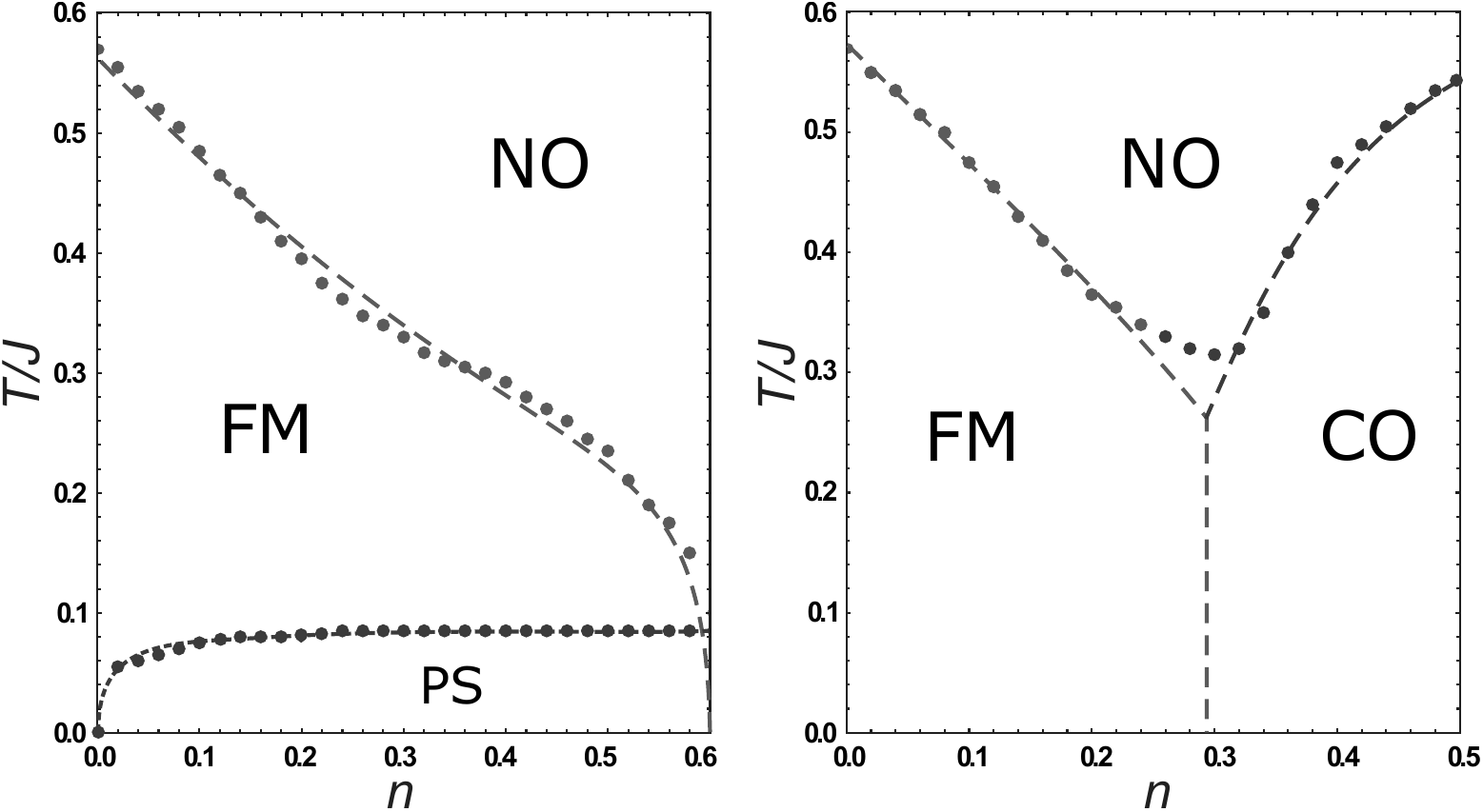} 
   \caption{\small{Phase diagrams for the system with Hamiltonian (\ref{MainHamiltonian}): (a) $V/\tilde{J} = 0.4$ (strong coupling) and (b) $V/\tilde{J} = 4$ (weak coupling). The temperatures corresponding to maximum heat capacity values, as calculated by a Monte Carlo method, are shown with filled
circles. Dotted lines schematically indicate boundaries of disordered (NO), ferromagnetic (FM), and charge-ordered (CO)
phases, as well as a phase separation region (PS).}}
   \label{fig:PhaseDiagram}
\end{figure}

The system described by Hamiltonian (\ref{MainHamiltonian}) exhibits
two types of phase diagrams that are exemplified in
Fig. \ref{fig:PhaseDiagram}. An example with a strong coupling ($\tilde{J}>V$) is
shown in the left panel (Fig. \ref{fig:PhaseDiagram}a). As the temperature is
lowered, two consecutive phase transitions occur in
the system in the range of $0 < n < 0.6$. The first one is
a transition from a high-temperature disordered state
into ordered ferromagnetic (FM) one diluted with
randomly distributed charged impurities. At low temperatures, mobile charged impurities condense into
"drops". This means that in case of strong coupling
the dilute FM phase is unstable with respect to phase
separation (PS) when the FM matrix expels impurities
in order to minimize surface free energy associated
with them. Weak coupling, i.e., when $\tilde{J}<V$, is exemplified in the right panel (Fig. \ref{fig:PhaseDiagram}b). With the impurity
concentration increasing, the FM spin ordering transforms into charge ordering (CO) of nonmagnetic
impurities. Further, we limit our discussion to the
range of $0 < n < 0.5$, because for $n > 0.5$ there is no long-range order in a spin system with a fairly high
critical temperature.

The competition between magnetic and charge
orders can be implemented in systems subject to disproportionation or instability with respect to charge
transfer fluctuations \cite{Moskvin2013}, e.g., as in cuprates. Typically,
with concentration n of impurities increasing, the
magnetic ordering is broken, the critical temperature
is lowered, and the temperature corresponding to
maximum values for MCE parameters, such as isothermal variation of magnetic entropy and adiabatic
temperature variation, is lower as well. Changes in the
maximum MCE temperature related to changes in the
chemical composition have been well demonstrated in
\cite{Shen2009,Inishev2018,Zhang2019}. However, the phase state of the system
described by Hamiltonian (\ref{MainHamiltonian}) also depends on the
ratio between the constants of exchange and density–density interactions, as was shown for a similar spin–pseudo-spin model \cite{PanovJLTP2016,PanovJSNM2016,Panov2017,Panov2019}. In this work, we compare
the MCE parameters in cases of strong and weak coupling; we also explore the possibility of detecting frustration in ground state of the system with weak coupling using magnetic entropy data.

\section{Calculation procedure for the MCE parameters using Monte Carlo data}

The adiabatic temperature change $\Delta T_{ad}$ and isothermal magnetic entropy change $\Delta S_{M}$ are the key parameters characterizing MCE. $\Delta S_{M}$ is calculated by
the equation:

\begin{equation}\label{DeltaEntropy1}
 \Delta S_{M} (T,h) = \int\limits_{0}^T \dfrac{C(T',h_m)-C(T',0)}{T'}dT',
\end{equation}
where $C$ is a heat capacity. Maxwell relationship $(\partial S/ \partial h)_T = (\partial M/ \partial T)_h$ gives us another expression for
 $\Delta S_{M}$ and an equation for $\Delta T_{ad}$:

\begin{equation}\label{DeltaEntropy2}
 \Delta S_M (T,h_m) = \int\limits_0^{h_m} \left( \frac{\partial M(T,h)}{\partial T} \right)_h dh
\end{equation}

\begin{equation}\label{DeltaTemperature}
	\Delta T_{ad} (T,h_m) = \int\limits_0^{h_m} \frac{T}{C(T,h)} \left( \frac{\partial M(T,h)}{\partial T} \right)_h dh
	,	
\end{equation}
where $M$ is a magnetization and $h$ is an external magnetic field. In case of second-order phase transitions,
expressions  (\ref{DeltaEntropy1}) and (\ref{DeltaEntropy2}) must lead to identical results. 
Further, magnetic field and temperature are expressed
in energy units, and entropy is expressed in $k_B$ units.

To obtain temperature dependences of MCE
parameters for the system with Hamiltonian  (\ref{MainHamiltonian}), we
used the Metropolis algorithm within the classical
Monte Carlo method. Calculations were carried out
for a 64$\times$64 square lattice with periodic boundary
conditions. In our program, we ensured that the condition for a constant concentration of impurities, i.e., $nN = \sum_i P_{0i} = const$ в, is held, since the variation in the
state of an arbitrarily chosen pair of lattice sites $a$ and $b$ was performed while sum $P_{0a} + P_{0b}$ was held constant.

We implemented a variant of this algorithm for parallel computations using NVIDIA graphic accelerators, which enabled us to perform calculations for
64 copies of the system simultaneously.  For each system copy, we lowered the temperature, while keeping
the external magnetic field constant, from value $T_1/J = 1.0$, which is around twice the magnetic ordering temperature, to $T_0/J = 0.01$ at a step of $T/J = 0.01$. We then varied the external magnetic field from $h/J = 0$ to $h_m /J = 0.04$ at a step of $h/J = 0.001$. Modeling was performed for impurity concentration  $0<n<0.5$  at a step $n = 0.1$. For each value for $n$, $T$ and $h$ we
obtained average values for energy, specific heat $C$ and magnetization $M$. This allowed us to use the trapezium rule for discretization of integrals and a threepoint method for derivative $( \partial M / \partial T )_h$ in Eqs.(\ref{DeltaEntropy1})--(\ref{DeltaTemperature}).

\begin{figure}[h]
   \centering
   \includegraphics[width=0.95\linewidth]{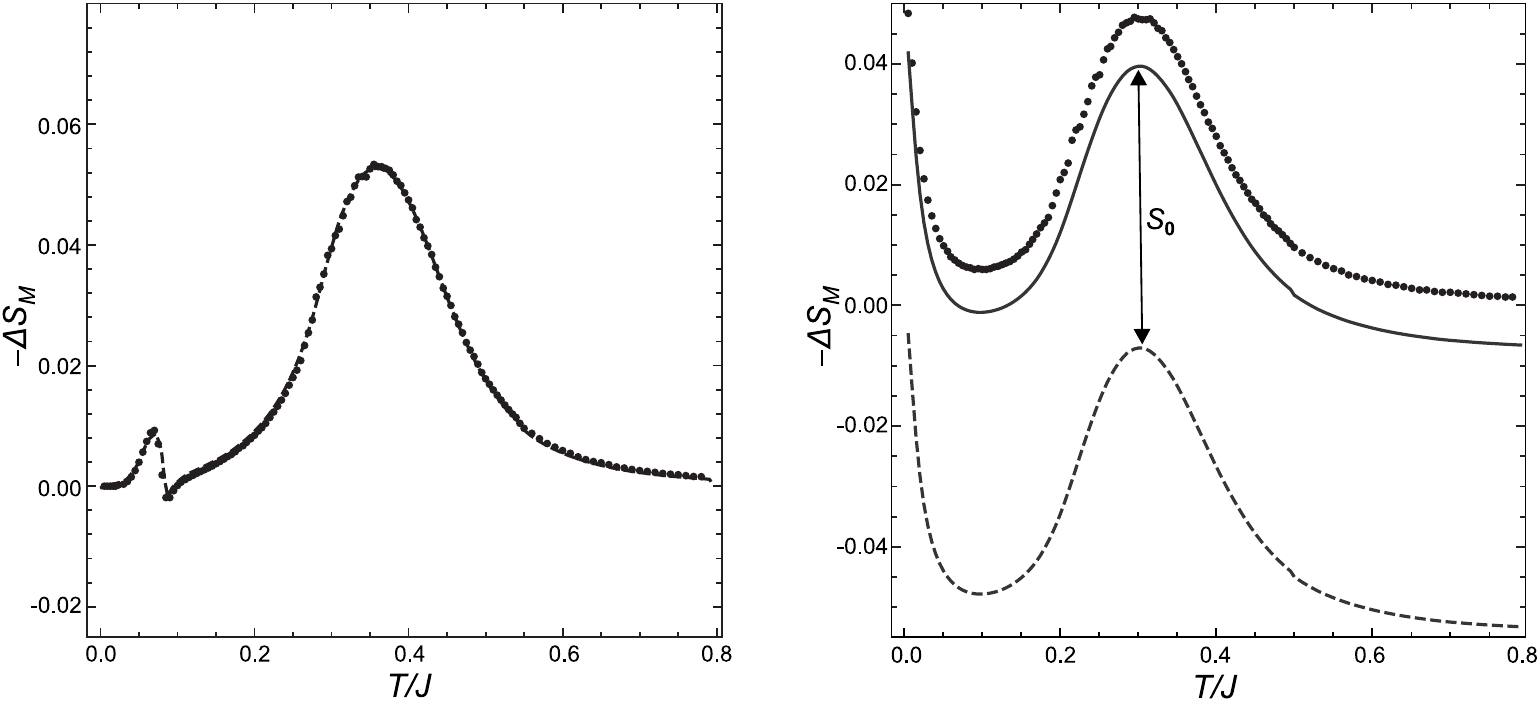}
   \caption{\small{ Magnetic entropy change for $n = 0.3$  in the case of (a) strong coupling ($V/\tilde{J} = 0.4$) and (b) weak coupling ($V/\tilde{J} = 4$). Results of calculations carried out using Eqs.(\ref{DeltaEntropy1}) and (\ref{DeltaEntropy2}) are shown with dotted lines and filled circles, respectively. A continuous
line pertaining to the case of weak coupling takes into account a contribution calculated by Eq.(\ref{EntropyChange}). The value corresponds to the
average concentration of paramagnetic clusters for $n = 0.3$ shown in Fig. 3. }}
   \label{fig:ZeroEntropy}
\end{figure}

With Eqs. (\ref{DeltaEntropy1}) and (\ref{DeltaEntropy2}) , we can calculate the variation of magnetic entropy using different Monte Carlo
data, i.e., the specific heat and magnetization. The
results of these calculations are in fairly good agree ment in case of strong coupling  ($\tilde{J} > V$) but at variance
in case of weak coupling ($\tilde{J} < V$). 
This is shown in Fig. \ref{fig:ZeroEntropy} for impurity concentration $n = 0.3$. 

\begin{figure}[h]
   \centering
   \includegraphics[width=0.95\linewidth]{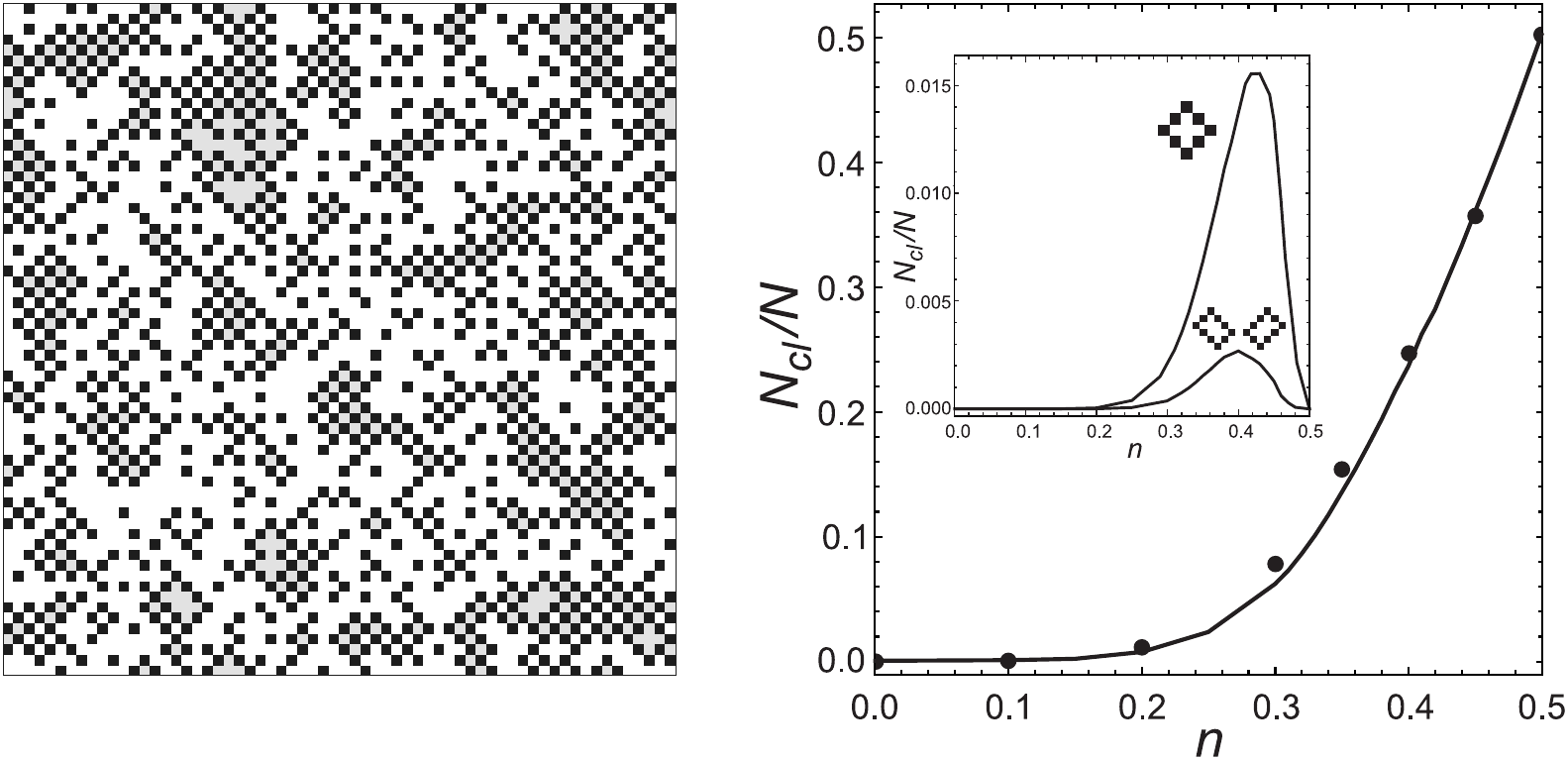}
   \caption{\small{ (a) Snapshot showing a configuration of the ground state for $n = 0.3$ and $V/\tilde{J} = 4$  (weak coupling). Lattice sites with $S_z = 1$ and $S_z = -1$ are represented by white and light-grey squares, respectively. Positions of nonmagnetic impurities ($S_z = 0$) are
designated with black squares.
(b) Average concentration of isolated paramagnetic clusters on a 64$\times$64 lattice. Inset: the results
for clusters comprised of 5 and 8 spins. The difference between maximum values for magnetic entropy variation  $\Delta S_M$, as calculated using Eqs.(\ref{DeltaEntropy1}) and (\ref{DeltaEntropy2})  and expressed in the units of average concentration of clusters (Eq.(\ref{ZeroEntropy})), is shown with filled circles.}}
   \label{fig:Clusters}
\end{figure}

In contrast to the case of strong coupling, with a
strong density–density interaction between impurities ($V > \tilde{J}$) , the value of paired distribution function for
the nearest neighbors turns 0 at a sufficiently low temperature and leads to charge ordering for $n\geq0.3$. 
A
picture characteristic of such charge ordering is shown
in Fig.\ref{fig:Clusters}. 
As a result, isolated spin clusters surrounded
by nonmagnetic impurities are formed. These clusters
behave like paramagnetic centers and furnish additional contribution  $S_0$ to the system entropy at the zero
temperature. The simplest estimation of this contribution can be based on the equation

\begin{equation}\label{ZeroEntropy}
 S_0 = N_{cl} \ln 2,
\end{equation}
where $N_{cl}$ is the number of isolated spin clusters. To
determine average number of clusters on a 64$\times$64, lattice, we used a separate program that generated random distributions of charged impurities having a minimal energy. The results of calculations are presented
in Fig.\ref{fig:Clusters}. Considering the contribution given by
Eq.(\ref{ZeroEntropy}), formula (\ref{DeltaEntropy1})  reads

\begin{equation}\label{EntropyChange}
  \Delta S_{M} (T,h) = S_0 + \int\limits_{T_{min}}^T \dfrac{C(T',h_m)-C(T',0)}{T'}dT',
\end{equation}
where $T_{min}$ is the minimum temperature used in
Monte Carlo simulations. As is shown in Fig.\ref{fig:ZeroEntropy}, this
improved consistency with $\Delta S_M$, M dependence that was
calculated using Eq.(\ref{DeltaEntropy2}). Thus, differences in results of
magnetic entropy  $\Delta S_M$ measurements that were
obtained through Eqs.(\ref{DeltaEntropy1}) and (\ref{DeltaEntropy2}) may point to some
hidden frustrations in the system. 

\section{Results}

The dependence of isolated cluster concentration
and the difference between maximum values for magnetic entropy change $\Delta S_M$ as calculated by Eqs. (\ref{DeltaEntropy1}) and (\ref{DeltaEntropy2}) and expressed in the units of average concentration, are presented in Fig.\ref{fig:Clusters}. We highlight a good
agreement between the cluster concentration and the
frustration data for the ground state of the system
obtained from MCE. Partial contribution of clusters
consisting of 5 and 8 spins are shown in the inset to this
figure. The maximum values for these contributions
are an order of magnitude lower than the total concentration of isolated paramagnetic clusters for a given $n$, which draws us to a conclusion that smallest clusters,
including 1 spin, make a decisive contribution to the
ground state entropy. In this case, the concentration of
smallest paramagnetic clusters increases monotonically with increasing $n$ reaching maximum $N_{cl}/N = 0.5$ at $n=0.5$.

\begin{figure}[h]
   \centering
   \includegraphics[width=0.75\linewidth]{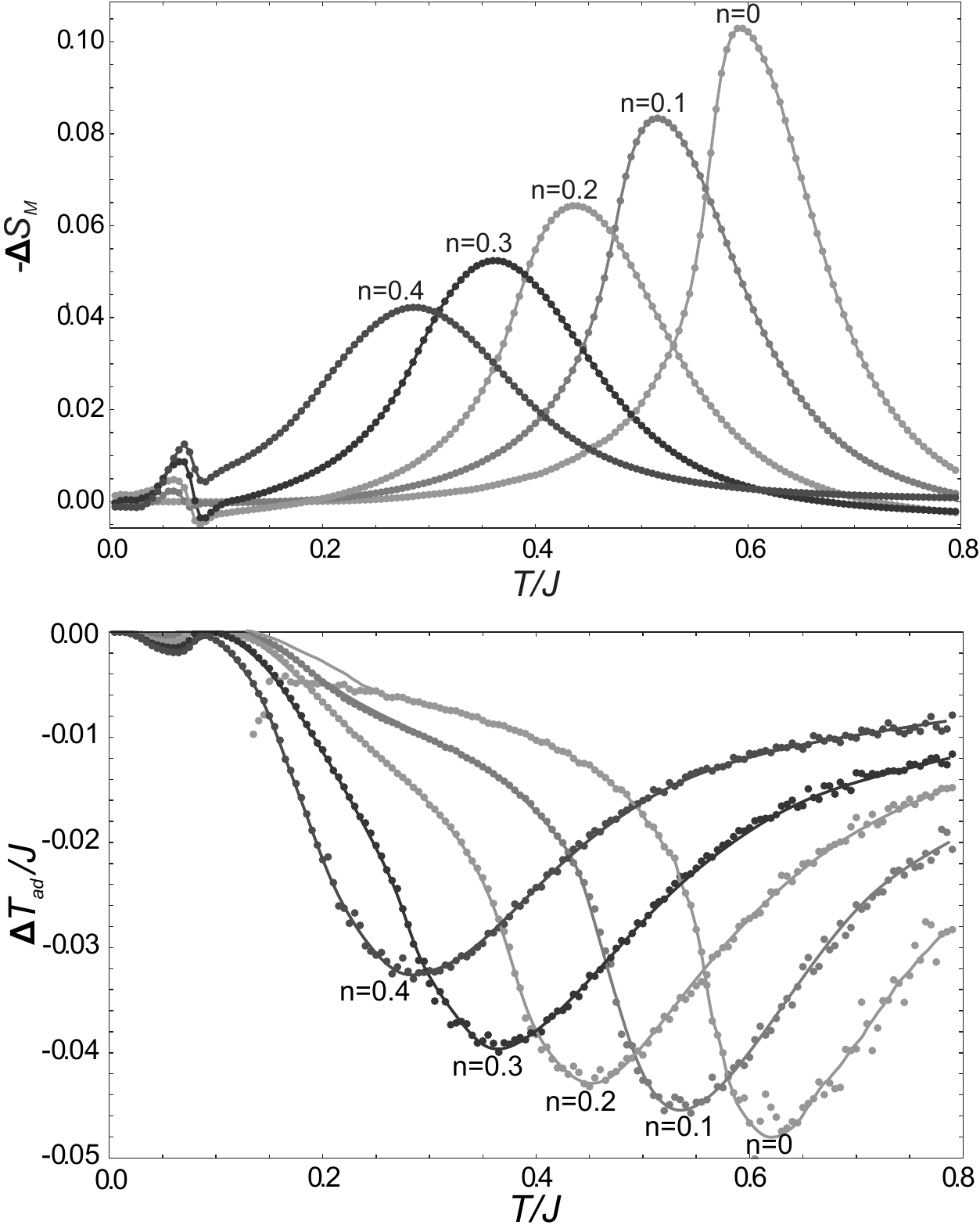}
   \caption{\small{ Magnetic entropy change $\Delta S_M$ and adiabatic temperature change $\Delta T_{ad}$ for concentrations $n$ of nonmagnetic impurities in
the range of $0–0.4$ in case of strong coupling ($V/\tilde{J} = 0.4$).}}
   \label{fig:StrongExchange}
\end{figure}

\begin{figure}[h]
   \centering
   \includegraphics[width=0.75\linewidth]{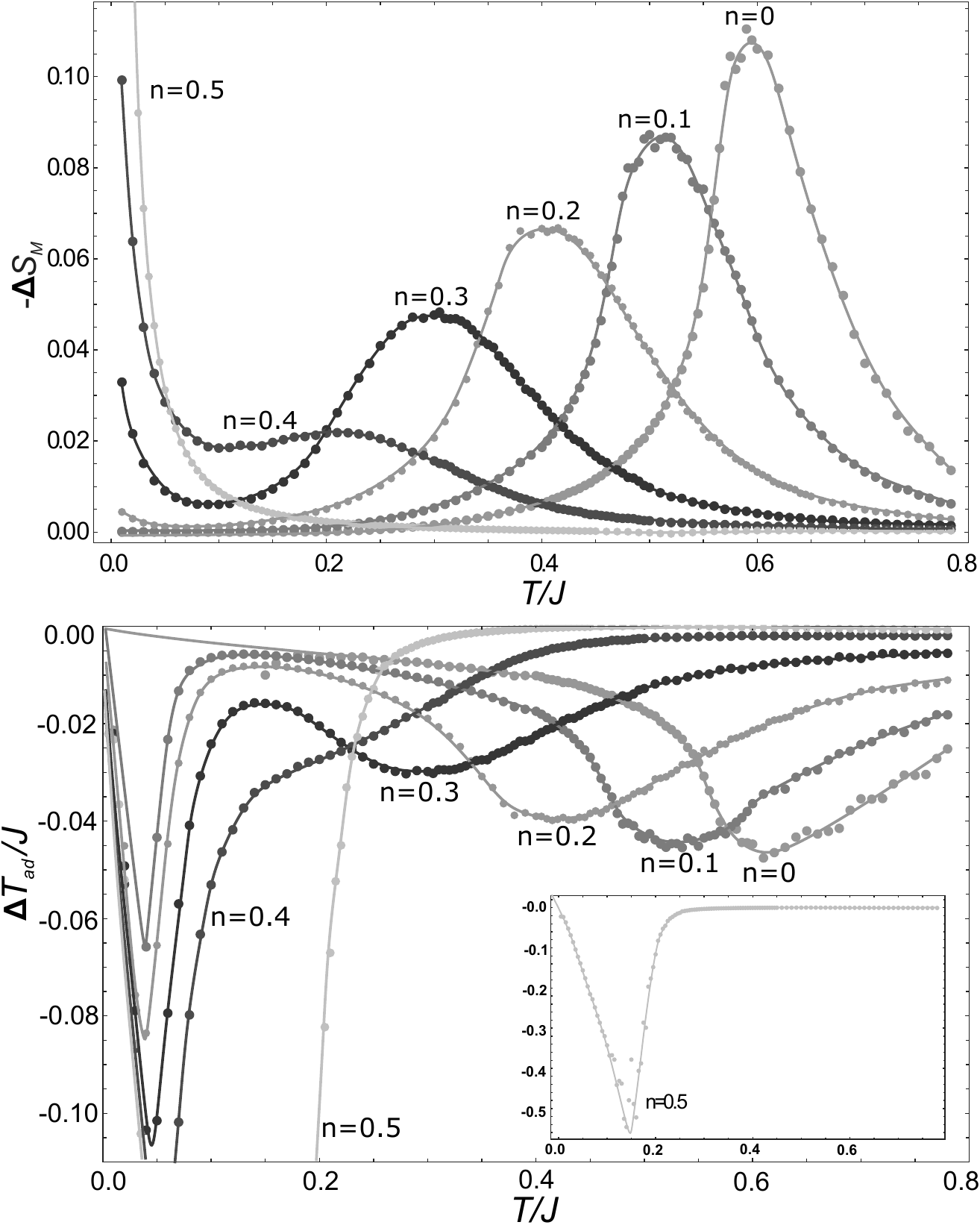}
   \caption{\small{ Magnetic entropy change $\Delta S_M$ and adiabatic entropy change $\Delta T_{ad}$ for concentrations $n$ of nonmagnetic impurities in the
range of $0–0.5$ in case of weak coupling ($V/\tilde{J} = 4$). Inset: results for $n=0.5$.}}
   \label{fig:WeakExchange}
\end{figure}

Using a discrete approximation for Eqs. (\ref{DeltaEntropy2}) and (\ref{DeltaTemperature}) we calculated isothermal magnetic entropy
change $\Delta S_M$ and adiabatic temperature change $\Delta T_{ad}$ on
the basis of Monte Carlo data. The results for $\Delta S_M$ and $\Delta T_{ad}$ in the case of strong coupling, i.e., $V/\tilde{J} = 0.4$, for $n = 0.0, 0.1, 0.2, 0.3, 0.4$ are shown in Fig.\ref{fig:StrongExchange}. The
temperatures corresponding to maximum values for
the two parameters follow approximately the variation
of magnetic ordering temperature with $n$. The peak
width grows, while their height diminishes as $n$ increases, because of impurity-induced smearing of
the phase FM transition. At low temperatures, dependences $\Delta S_M$ and $\Delta T_{ad}$ feature an additional peak
caused by phase separation.
$\Delta S_M$ and $\Delta T_{ad}$ in the case of weak coupling, i.e., $V/\tilde{J} = 4$, for $n = 0.0, 0.1, 0.2, 0.3, 0.4, 0.5$ are shown
in Fig.\ref{fig:WeakExchange}. For $n = 0.0$ and $n = 0.1$, these results are nearly
identical to those for strong coupling. However, qualitative differences are seen at $n > 0.1$ , especially at low
temperatures. As was discussed above, charge ordering
in the impurity subsystem leads to the formation of
isolated spin clusters. This causes a paramagnetic
response of the MCE parameters that gives maximum
value $\Delta T_{ad}/J \approx 0.55$ for $n = 0.5$ and $T/J = 0.14$.

\section{Conclusions}

We performed Monte Carlo simulations of a twodimensional Ising system containing a fixed concentration of mobile nonmagnetic charged impurities,
which enabled us to construct temperature dependences of the MCE parameters. It was shown that in
case of weak coupling, for nonmagnetic impurity concentration $n>0.1$ , isolated spin clusters are formed in
the ground state of the system. This leads to nonzero
entropy of the system’s ground state. The feasibility to
detect a frustration in the ground state is demonstrated
using the magnetic entropy variation data. The MCE
parameters were calculated for the cases of strong and
weak coupling. 

\section*{Funding}

The work was supported within the Competitiveness
Enhancement Program of the Ural Federal University
(Russian Federation Government Act 211, agreement no.
02.A03.21.0006), by the Ministry of Education and Science, Russian Federation (project FEUZ$-2020-0054$),  and
by the Russian Foundation for Basic Research (research
project no. $18-32-00837\backslash 18$).

\section*{Conflict of interest}

The authors declare that they have no conflicts of interest.
\newpage


\end{document}